# The structural and optical properties of $In_xGa_{1-x}N$/GaN epilayers grown on a miscut sapphire substrate

I.A. Ajia[1], S.M.C. Miranda[2,*], N. Franco[2], E. Alves[2], K. Lorenz[2], K.P. O'Donnell[3], and I.S. Roqan[1,†]

[1] King Abdullah University of Science and Technology, Physical Sciences and Engineering Division, Saudi Arabia

[2] IPFN, Instituto Superior Técnico, Campus Tecnológico e Nuclear, Bobadela, Portugal

[3] Department of Physics, SUPA, University of Strathclyde, Glasgow, United Kingdom

We report on structural and optical properties of InGaN/GaN thin films, with a 0.46° misalignment between the surface and the (0001) plane, which were grown by metal-organic chemical vapor deposition (MOCVD) on 0.34° miscut sapphire substrates. X-ray diffraction and X-ray reflectivity were used to precisely measure the degree of miscut. Reciprocal space mapping was employed to determine the lattice parameters and strain state of the InGaN layers. Rutherford backscattering spectrometry with channeling was employed to measure their composition and crystalline quality with depth resolution. No strain anisotropy was observed. Polarization-dependent photoluminescence spectroscopy was carried out to examine the effect of the miscut on the bandedge emission of the epilayer.

* Current address: Instituut voor Kern- en Stralingsfysica, KU Leuven, 3001 Leuven, Belgium
† The corresponding author: iman.roqan@kaust.edu.sa

## 1. Introduction

InGaN/GaN based light emitting diodes (LEDs) and laser diodes are pervasive due to their high luminescence efficiency. The effects of the strain induced into the heterostructure based on InGaN materials on their optical properties is subject of extensive research.[1-5] Obtaining high quality surface morphology for a GaN or InGaN template grown on sapphire ($Al_2O_3$) is significant to improving the performance of III-nitride quantum wells (QWs)-based LEDs.[6-8] Therefore, for the growth of III-nitrides on a foreign substrate, a substrate miscut at a certain angle has been introduced to enhance the surface morphology of the overgrown layers.[9, 10] As the *c*-parameter of the $Al_2O_3$ is different from that of GaN, the small difference in the length of the *c*-parameter of the substrate and that of the epilayer depends on the epilayer *c*-plane tilt angle with respect to the substrate lattice plane.[9] Therefore, the tilt angle of the miscut substrate should be optimal in order to obtain the desired surface. Kryśko *et al.* found that the InGaN epilayer tilt was introduced due to $Al_2O_3$ substrate miscutting.[8] In addition, miscutting also affected the In content.[7] Shojiki et al. reported that the In content increases as the miscut angle of the *c*-plane $Al_2O_3$ substrate around the a-axis increases.[7] The effect of In content on the quality of the luminescence properties of InGaN/GaN heterostructure and quantum wells grown on substrate miscut at different angle has also been investigated.[11, 12] There is, however, no systematic study on the effect of the optimum substrate miscut angle on the material properties for different InN contents. Moreover, whether this miscut affects the strain isotropicity (of the InGaN layers and their structural quality that can be used as a template for QW growth remains insufficiently investigated. Strain anisotropy affects the energy isotropicity of the InGaN bandedge. The bandedge energy anisotropy occurs for III-nitrides grown in semi-polar and non-polar directions by breaking the symmetry of the valence band (VB) wavefunction and induces the separation of the VB into a distinct maxima. This, in turn, causes a dependency of the near band edge (NBE) emission on the polarization angle of the excitation source.[13, 14] This energy anisotropy occurs when InGaN experiences an anisotropic compressive strain in growth directions that are not parallel to the *c* axis. Such VB separation occurs at the Γ (k=0) point resembling the $|X\rangle$ ([11-20]), $|Y\rangle$ ([1-100]) and $|Z\rangle$ ([0001]) components, corresponding to the heavy hole (HH), light hole (LH), and crystal-field split-off hole (CH), respectively.[15, 16]

In this work, we investigate the structural and optical properties of InGaN epilayers grown on a GaN buffer layer on a miscut polar $Al_2O_3$ substrate with the aim of obtaining a smooth surface morphology. Using X-ray diffraction and X-ray reflectivity, we establish a method of measuring the slightly misaligned InGaN from the *c* direction and found that the miscut is initiated from the $Al_2O_3$ substrate and causes a misorientation of the *c*-planes of the GaN buffer layer with the surface. Thus, we can show the effect of the optimum miscut on the epilayer strain and the related optical properties.

## 2. Material and methods

High quality thin $In_xGa_{1-x}N$/GaN bilayers (produced commercially by TopGaN[17]) were grown by MOCVD on *c*-$Al_2O_3$ (0001) substrates. The targeted indium nitride fractions are in the 0.3% ≤ x ≤ 14% range as the growth temperatures declines from 920 $^oC$ to 790 $^oC$, as shown in Table 1. To determine the material quality, Rutherford backscattering spectrometry / channeling (RBS/C) measurements were performed using 2 MeV $He^+$ ions and a Si surface barrier detector mounted in IBM geometry at a backscattering angle of 140$^o$. 'Random' spectra were taken at 5$^o$ and 20$^o$ tilt angles and were fitted simultaneously using the NDF code.[18] Furthermore, aligned spectra were acquired along the *c*-axis to assess the crystalline quality of the layers. To determine the lattice parameters, strain state, miscut of the $Al_2O_3$ substrates and the misalignment between different sample layers, high resolution X-ray diffraction (XRD) and X-ray reflectivity (XRR) measurements were carried out using monochromated $CuK_{\alpha1}$ radiation on a Bruker-AXS D8Discover system, employing a Göbel mirror and an asymmetric 2-bounce Ge(220) monochromator in the primary beam. XRD rocking curves (RC) were measured using the open scintillation detector. XRR as well as reciprocal space maps (RSM) around the 10−15 reciprocal lattice point were acquired using a 0.1 mm slit placed in front of the detector in the secondary beam. Photoluminescence (PL) measurements were carried out using a monochromated 1000 W Xe arc lamp at 10 K. The PL polarization and temperature-dependent PL measurements were carried out using a vertically polarized He-Cd laser operating at 325 nm. The samples were mounted in a closed-cycle helium cryostat. The excitation light was incident at an angle of ~60$^o$ to the surface normal for the observation of polarization effects.[19] For polarization angle-dependent measurements, the laser beam was first expanded and depolarized. The excitation light was subsequently repolarized with a Glan-Thomson prism mounted on a 360$^o$ adjustable stage with fine angular adjustment. The light was then focused on the samples in the cryostat using a plano-convex lens.

## 3. Results and discussions

The InN compositions and the layer thicknesses of four samples, labeled A-D, were measured using the random RBS spectra shown in Fig. 1. The InN composition profile is shown to be uniform throughout the layer thickness for Sample A, B, and C. However, Sample D (with high average InN content) shows enhanced InN incorporation near the surface compared to that at the interface; such behavior is common for InN-rich layers.[20] Table 1 summarizes the RBS results.

To evaluate the crystal quality of the samples, RBS/C was carried out. Crystal disorder and crystal quality can be quantified by the minimum yield, $\chi_{min}$,[21] denoting the ratio of backscattering yields from the aligned spectrum to that from the random spectrum. RBS/C spectra in Fig. 1 show that Sample A, B, and C have high crystal quality, similar to that of state-of-the-art GaN binary crystal[22] with $\chi_{min}$ = 2, 2.5 and 7%, respectively (Table 1). However, Sample D shows a significantly poorer $\chi_{min}$ of 70%, indicating low crystal quality, as shown clearly in Fig. 1.

GaN buffer layer *a* and *c* lattice parameters were determined using Bond's methods, yielding values typical of GaN grown on $Al_2O_3$ (*a* = 3.184(1), *c* = 5.189(1)).[23] After the samples were accurately aligned using the GaN diffraction peak, the lattice parameters of the $In_xGa_{1-x}N$ epilayers were extracted from the asymmetric (10−15) RSMs shown in Fig. 2. To calculate the composition of the layers from these values (Table 1), biaxial strain was taken into account using Poisson's equation. The strain-free $a_0$ and $c_0$ lattice parameters of GaN and InN are taken from Deguchi *et al.*,[24] while their respective stiffness coefficient values $C_{13}$ and $C_{33}$ are adopted from extant work.[25] Only the GaN diffraction signal was obtained in Sample A, in which the InGaN peak overlaps with the intense signal from the GaN buffer layer. (Note that the second peak above that corresponding to the GaN buffer layer is due to the $K_{\alpha 2}$ line, which is not completely suppressed by the monochromator.) The RSMs from Sample B and C show that the $In_xGa_{1-x}N$ epilayers are pseudomorphically strained to the GaN buffer, as their Qx positions correspond to that of GaN. In addition, the high symmetry of the $In_xGa_{1-x}N$ reflection signal from these samples, as shown in Fig. 2(b) and (c), signifies high strain homogeneity across the films. The small broadening of the peak also suggests high crystallinity, in agreement with the RBS/C results. As expected, the *c* parameter value increases with InN content (Table 2). In contrast, Sample D is found to be partially relaxed, which produces a very weak and broad peak (figure not shown). Due to the poor crystal quality of Sample D revealed by RBS/C and XRD, this sample was excluded from the optical polarization measurements described below.

To determine the miscut angles, X-ray reflectivity rocking curves (RCs) were first measured as a function of the azimuthal angle Φ at fixed 2θ = 0.5°. These measurements are sensitive to the sample surface position. Then, XRD RCs of the 006 reflection of $Al_2O_3$ and the 002 reflection of GaN and InGaN (if distinguishable from the GaN peak) were acquired, again as a function of Φ. The RC centers ($\theta_{max}$) were then plotted as a function of Φ (Fig. 3) and the function given below was fitted to the experimental results:

$$\theta_{\max} = B + A\sin(\Phi + \Psi), \qquad (1)$$

where *B* corresponds to the Bragg angle of the chosen reflection—0.25° in the case of X-ray reflectivity (XRR)—and the contribution of an inherent misalignment between the beam, the goniometer and the sample holder, and *A* is the amplitude that is dependent on the angle between the surface in the case of XRR (diffraction planes in the case of XRD) and the sample holder plane on which the sample is mounted. The fitting parameters A and Ψ describe the vector corresponding to the normal to the measured plane from which the relative orientation of the different layers can be derived. Assuming that the sample surface is parallel to that of the $Al_2O_3$ substrates, for all samples, $Al_2O_3$ exhibits a slight misalignment of ~0.34°, as shown in Table 3. In addition, the GaN 002 planes are tilted by ~0.13° from the 006 $Al_2O_3$ planes. The sum of these two values is in good agreement with the directly measured angle between the surface and (0001) planes of GaN. The GaN and $In_xGa_{1-x}N$ layers are parallel to each other within experimental accuracy. The miscut analysis results are summarized in Table 3. However, no strain anisotropy is observed for these samples.

Fig. 4(a) shows PL spectra pertaining to samples A−C at 12 K. Sample A, with 0.4% InN, exhibits only a spectral response typical of GaN, dominated by NBE emission at 357 nm. Sample B, with 8.9% InN, has a prominent InGaN peak at 404 nm, whereas Sample C (15.8% InN) has a dominant peak at 442 nm attributed to the InGaN NBE emission. In addition, spectra of both Sample B and C include a small component of GaN NBE emission at 357 nm.

Only Sample B and C were subjected to the polarization PL measurements to investigate if there is an energy anisotropy, since fully compressive strain is a prerequisite for the occurrence of this anisotropy. As previously discussed, as Sample D is relaxed due to high InN content, it was not suitable for the polarization study. It has to be considered that anisotropic polarization response can be induced for certain incident angles of the polarized excitation laser.[19] Therefore, we carried out such polarization measurements at low temperature, as shown in Fig. 4(b) and (c). We observed an energy shift of the NBE peak in the PL spectra of Sample B and C for different polarized excitation light. In both samples, there is a small, but definite, blue-shift as the polarization angle of the excitation is switched from $\parallel c$ to $\perp c$. On the other hand, we carried out the polarization measurement at the same incident angle on pure GaN grown on $c$-$Al_2O_3$ (not shown) and did not observe any signs of NBE peak energy shift as the polarized excitation laser angle changed. In Fig. 5(a), we show the angular response of Sample B PL spectra taken at the low temperature of 7 K. The degree of polarization, ρ, is determined using the equation:

$$\rho = \frac{I_\perp - I_\parallel}{I_\perp + I_\parallel}, \qquad (2)$$

where $I_\perp$ is the PL intensity of the sample excited with the laser polarized perpendicularly to the $c$-axis, and $I_\parallel$ is the component of the PL intensity with polarization parallel to the $c$-axis. The maximum peak shift occurs at 3.08 eV, as shown in Fig. 5(a), corresponding to 31% polarization.

The inset gives the angular dependence of the PL peak energy at 7 K and 40 K, indicating absence of temperature dependence. At these temperatures, as the polarization angle is changed from 180$^o$ (∥*c*) to 270$^o$ (⊥*c*), a gradual blue-shift in the bandedge peak is observed. At 7 K (40 K), the peak energy value rises from 3.06 eV (3.05 eV) to 3.07 eV (3.06 eV) with the increase in the polarization angle. Although polarization selection rules necessitate that the peak of the ⊥*c* component occupies the lowest energy,[26] they were contradicted in the case of our $In_xGa_{1-x}N$ thin epilayers. With polarization parallel to the basal axes (∥*c*), the samples exhibited higher peak energy than that obtained for ∥*c*. However, this argument is insufficient to prove the anisotropy, as investigating the cause of such energy shift by measuring polarization dependence on $In_xGa_{1-x}N$ layers with growth direction along *c*-axis is not an easy process. Hence, for further investigation, we carried out polarization measurements on the emitted light as the function of the angle of the incident laser with respect to the *c*-axis. No energy anisotropy was observed for the emitted light. Therefore, such energy shifts of the different polarized light cannot be due to strain anisotropy and the VB separation. Thus, the blue-shift of the peak wavelength can be due to the band-tail filling effect by carriers[27] that is observed on the m-plane. Such band filling effects can be due to fluctuations of the bandgap of inhomogeneous In distribution across the sample that creates localized states.[28] In this case, when the laser intensity changes due to different polarization degrees, the filling of the low energy band-tail states occurs by the carriers, causing such blue-shift in the NBE peak. Therefore, as the miscut angle is small, it cannot affect the isotropy of the bandgap or the strain.

## 4. Conclusions

In summary, we have carried out structural and optical analyses on slightly miscut InGaN thin film samples. Structural analyses showed that the miscut InGaN thin epilayer was pseudomorphically strained for low In content samples with a state-of-the-art GaN-like quality.

We found that polarization-dependent PL of excited light showed blue-shift energy with different degrees of light polarization, which may have been induced by the slight misalignment of the InGaN layers. The polarization measurements of the emitted light did not exhibit an energy shift, indicating that energy anisotropy did not occur for InGaN slightly tilted from the *c*-axis. Thus, we conclude that the blue-shift of the NBE emission with polarization is due to the band filling effect.

**Figure captions**

Fig.1: Measured and simulated **(**solid lines**)** RBS/C spectra of InGaN layers grown at (a) 920$^o$C (sample A) (b) 820 $^o$C (sample B) (c) 780 $^o$C (Sample C) and (d) 710 $^o$C (sample D). Random spectra were acquired with a tilt of 5$^o$ and 20$^o$ between the surface normal and the incoming beam. The $\chi_{min}$ values have been calculated from the 5$^o$ and the aligned spectra which have been acquired using the same integrated charge.

Fig.2: XRD RSM around the 10-15 reciprocal lattice point for samples A, B and C.

Fig. 3: XRR and XRD rocking curve peak positions (symbols) and fits using equation1 (solid lines) of sample C, measured as a function of the azimuthal angle (Φ). (a) Sample surface (XRR), (b) $Al_2O_3$ (0001) plane (XRD), (c) GaN (0001) plane (XRD), and (d) InGaN (0001) plane (XRD). See text for details.

Fig. 4: (a) Low temperature PL spectra for the samples taken at 12 K, using 340 nm monochromated Xe arc lamp. (b) and (c): Low temperature PL spectra showing energy blue-shift in the pseudomorphic InGaN films for samples B and C respectively.

Fig. 5: (a) Solid lines are PL spectra of polarized intensities of sample B at angles parallel and perpendicular to the *c*-axis. The dotted line is the degree of polarization for the sample. (b) Parallel and perpendicular polarization intensities with their multiple Gaussian components at 3.051 eV and 3.072 eV, respectively.

| Sample | Nominal Growth Temp (ºC) | $\chi_{min}$ (%) | InN content (%) (RBS) | InN content (%) (XRD) | Thickness (nm) (RBS) |
|---|---|---|---|---|---|
| A | 920 | 2.0 | 0.4 | n.a. | 40 |
| B | 820 | 2.5 | 8.9 | 9.0 | 45 |
| C | 780 | 7.0 | 15.8 | 15.9 | 50 |
| D | 710 | 70 | 30 (Surface) 27 (near the interface) | 23.9 | 29 (Surface) 26 (near the interface) |

Table 1. The growth temperature, InN contents (measured by RBS and XRD) and InGaN layer thickness measured by RBS.

| **Sample** | ***c* [Å] GaN** | ***a*[Å] GaN** | ***c* [Å] InGaN** | ***a*[Å] InGaN** | **$\varepsilon_{\parallel}$ [%]** | **$\varepsilon_{\perp}$ [%]** |
|---|---|---|---|---|---|---|
| A (only GaN detected) | 5.189(1) | 3.184(1) | n.a. | n.a. | n.a. | n.a. |
| B | 5.190(1) | 3.185(1) | 5.262(1) | 3.186(1) | -1.1 | 0.6 |
| C | 5.189(1) | 3.184(1) | 5.324(1) | 3.185(1) | -1.9 | 1.1 |

Table 2. GaN and InGaN lattice parameters and in-plane $\varepsilon_{\parallel}$ and out-of-plane $\varepsilon_{\perp}$ biaxial strain

| Sample | Angles between planes (°) | | | |
|---|---|---|---|---|
| | **Surf*$Al_2O_3$** | **Surf*GaN** | **$Al_2O_3$*GaN** | **$Al_2O_3$*InGaN** |
| B | 0.31 | 0.43 | 0.13 | n.a. |
| C | 0.36 | 0.48 | 0.13 | 0.14 |

Table 3. Miscut angles estimated by XRD measurements.

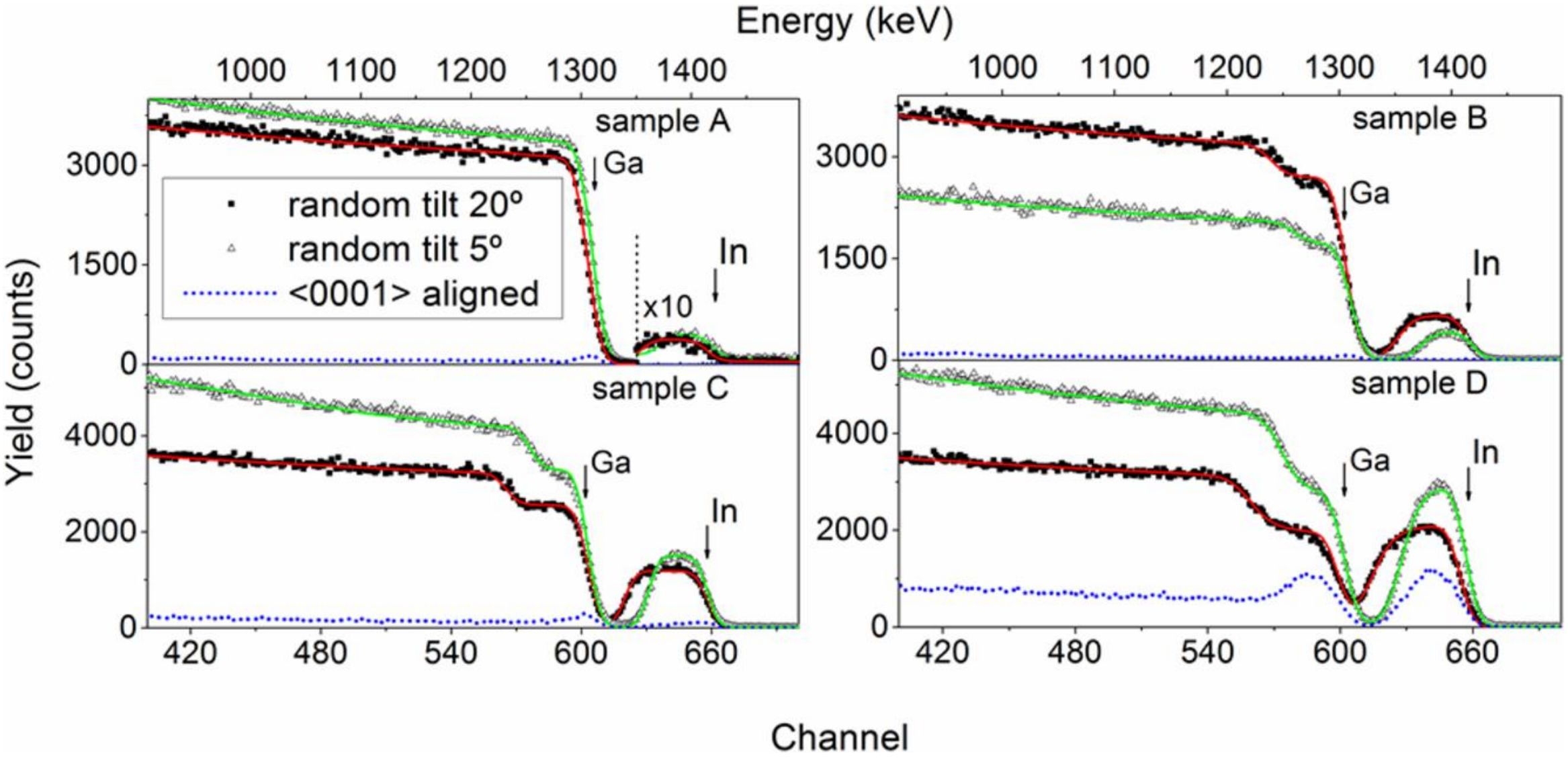

Fig. 1

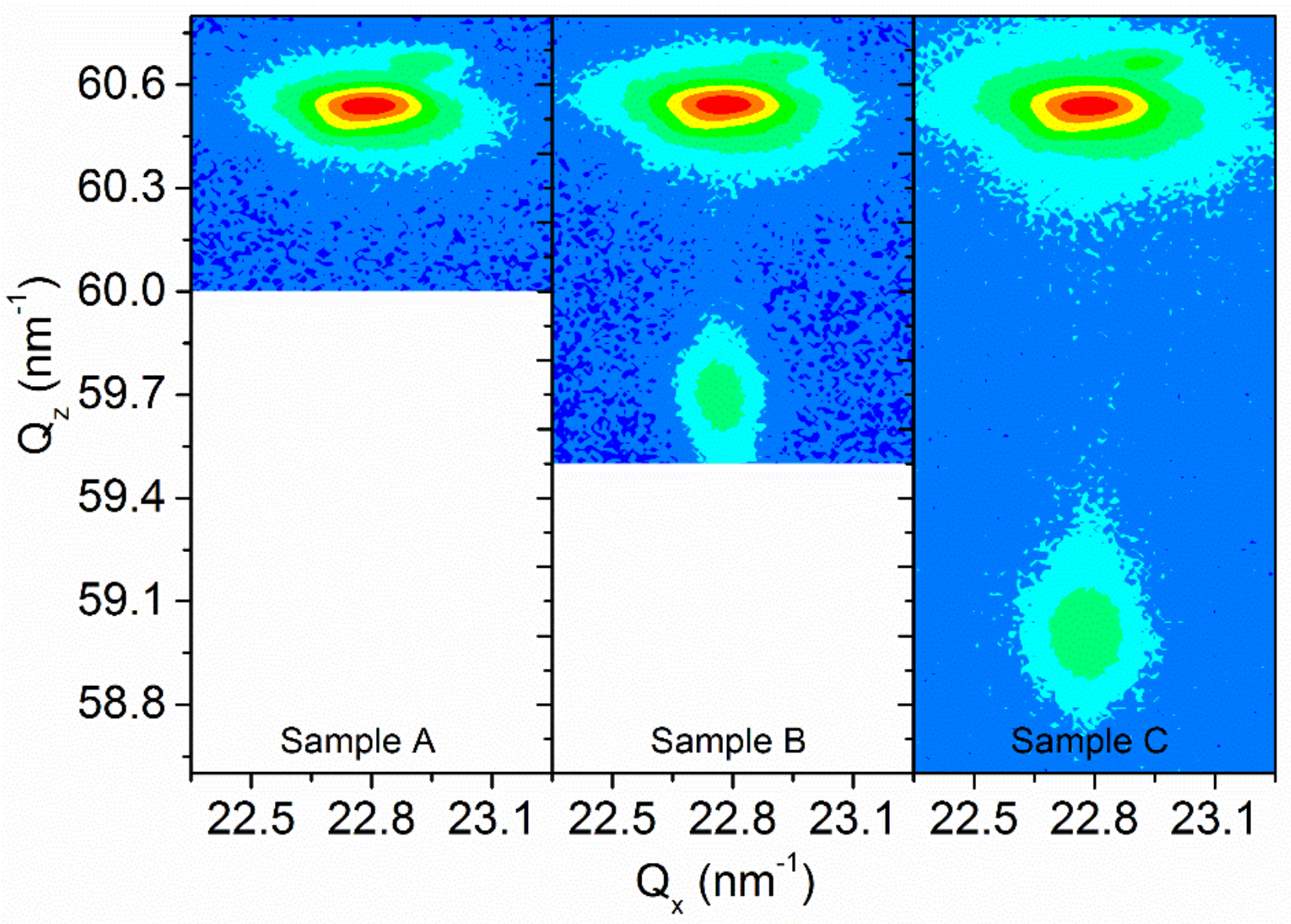

Fig. 2

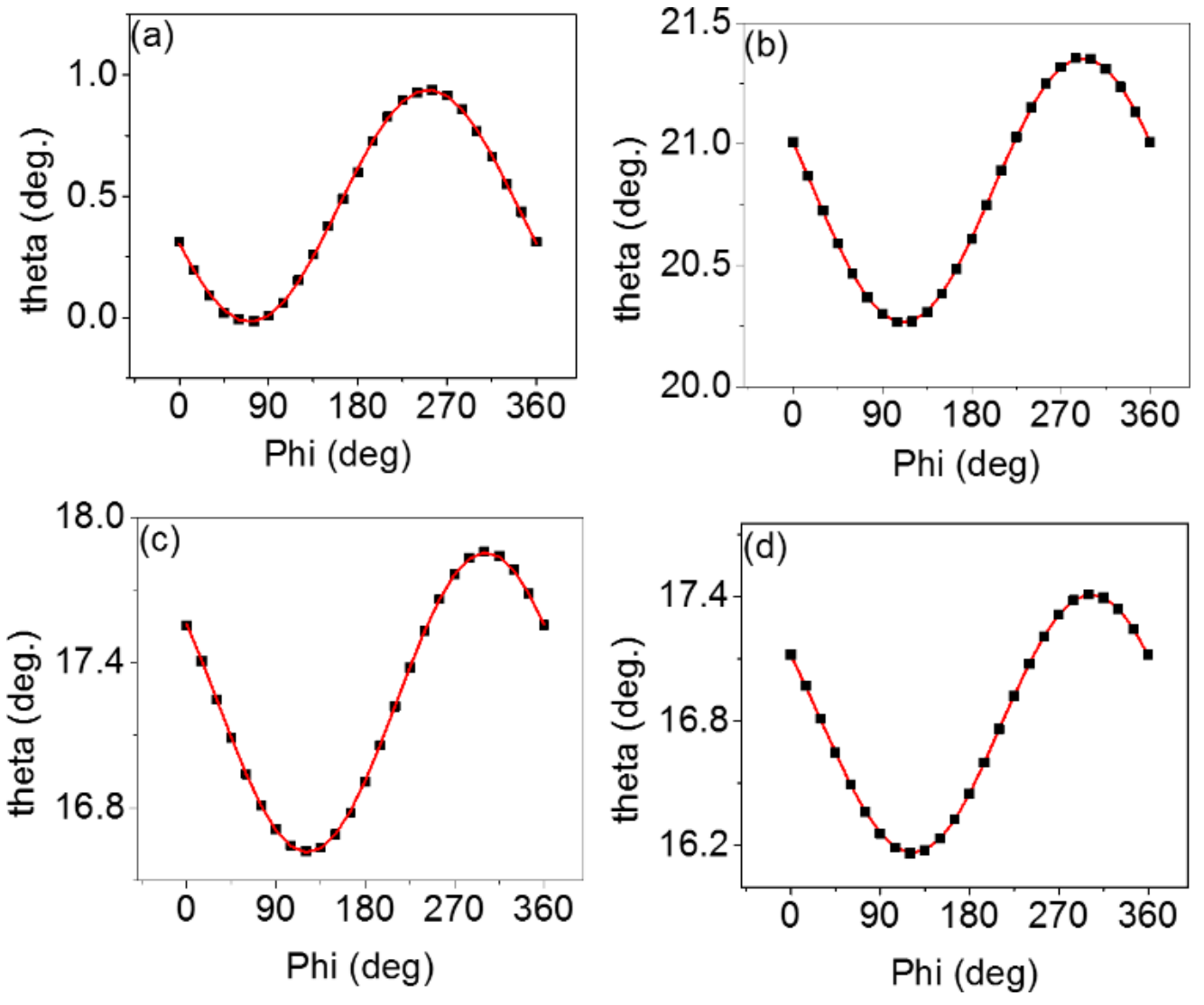

Fig. 3

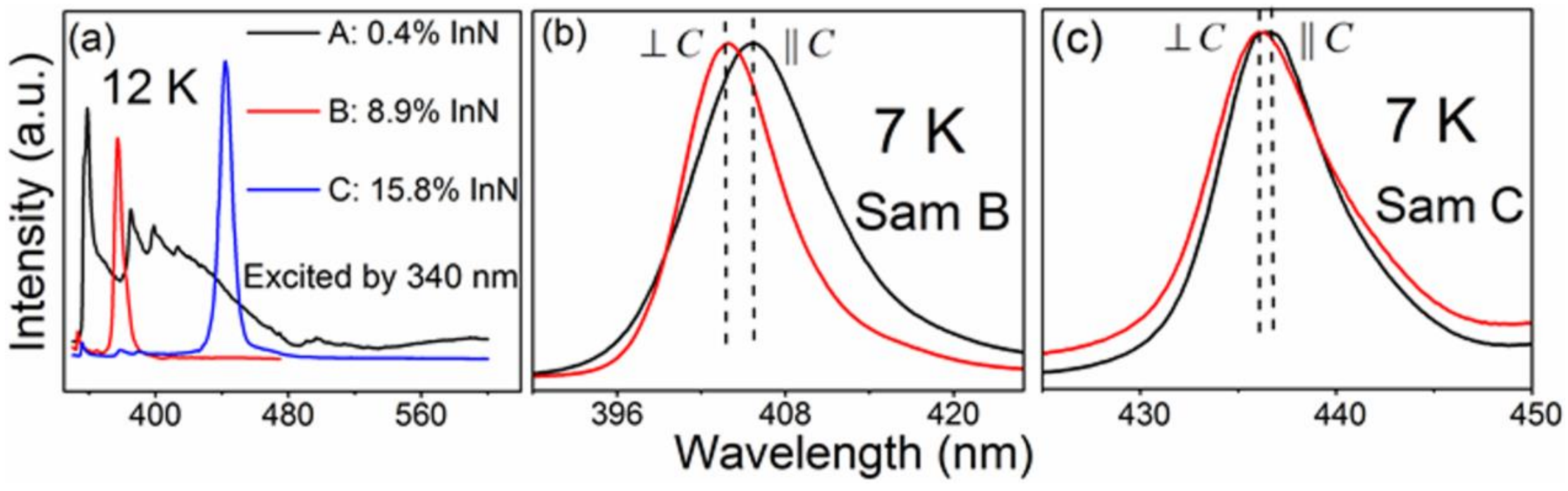

Fig. 4

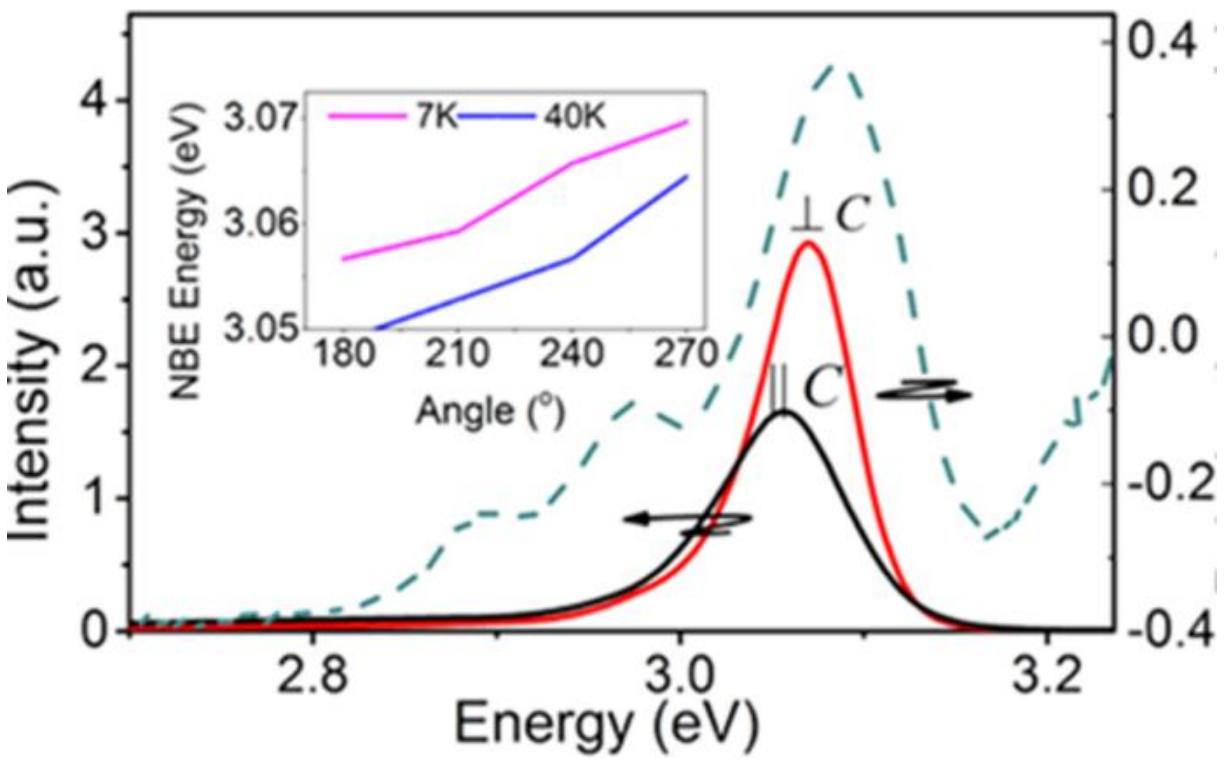

Fig. 5